\documentclass[twocolumn,showpacs,reprint,amsmath,amssymb,showkeys,aps,prb]{revtex4-1}

\usepackage{graphicx}
\usepackage{dcolumn}
\usepackage{bm}
\usepackage{bezier}
\usepackage{xcolor}
\usepackage[utf8]{inputenc}
\usepackage[T1]{fontenc}
\usepackage[english]{babel}

\begin{document}

\title{Full counting statistics of quantum phase slips}
\author{Andrew G. Semenov$^{1,3}$
and Andrei D. Zaikin$^{2,1}$
}
\affiliation{$^1$I.E.Tamm Department of Theoretical Physics, P.N.Lebedev
  Physical Institute, 119991 Moscow, Russia\\
$^2$Institute of Nanotechnology, Karlsruhe Institute of Technology (KIT), 76021 Karlsruhe, Germany\\
$^3$National Research University Higher School of Economics, 101000 Moscow, Russia}

\begin{abstract}
We work out a microscopic theory describing complete statistics of voltage fluctuations generated by quantum phase slips (QPS) in superconducting nanowires. We evaluate the cumulant generating function and demonstrate that shot noise of the voltage as well as the third and all higher voltage cumulants differ from zero only due to the presence of QPS. In the zero-frequency limit voltage fluctuations in superconducting nanowires are described by Poisson statistics just as in a number of other tunneling-like problems. However, at non-zero frequencies quantum voltage fluctuations in superconducting nanowires become much more complicated and are not anymore accounted for by Poisson statistics. In the case of short superconducting nanowires we explicitly evaluate all finite-frequency voltage cumulants and establish a non-trivial relation between these cumulants and the current-voltage characteristics of our system.
\end{abstract}

\maketitle

\section{Introduction}
Superconducting fluctuations play a prominent role in a reduced
dimension \cite{LV}. Such fluctuations become particularly pronounced in
quasi-one-dimensional superconductors \cite{AGZ} which properties
drastically differ from those of bulk systems. For instance, small
fluctuations of the superconducting phase are converted to sound-like
plasma modes \cite{ms,Bu} which can propagate along superconducting
nanowires forming an effective dissipative environment for electrons
inside the wire. Interaction with this environment yields smearing of
the gap singularity in the electron density of states and generates
non-vanishing tail of states at subgap energies for any non-zero temperature
\cite{RSZ,ALRSZ}.

In addition to small phase fluctuations, at low enough $T$ quasi-one-dimensional superconducting wires host another type of fluctuations called  quantum phase slips \cite{AGZ,ZGOZ,GZQPS,BT,Lau,Zgi08,AstNature} (QPS). In the course of a QPS event the superconducting order parameter temporarily drops down to zero at some point of the wire and, hence, the superconducting phase there becomes unrestricted. Later on the order parameter gets restored and its phase can change by $\pm 2\pi$ as compared to its initial value. In accordance with the Josephson relation such phase jumps yield voltage pulses. Breaking the symmetry between $+2\pi$ and $-2\pi$ pulses by applying a bias current one generates non-zero average voltage across the wire. Thus,  in the presence of QPS quasi-one-dimensional superconductors acquire non-zero resistance \cite{ZGOZ}. 

Note that at $T=0$ sufficiently thick wires demonstrate (almost) superconducting behaviour  meaning that their linear resistance tends to zero. In contrast, thinner wires turn insulating. This non-trivial behavior is fully controlled by quantum phase slips which can formally be viewed as logarithmically interacting vortices is space-time. It follows immediately that there exists a quantum phase transition corresponding to unbinding of QPS-anti-QPS pairs at some critical value of the wire thickness \cite{ZGOZ}. This superconductor-insulator transition (SIT)  belongs to the same universality class as the Berezinskii-Kosterlitz-Thouless phase transition in classical 2d systems.

Yet another fundamental property of the systems under consideration is the duality between the phase and the charge spaces \cite{PZ88,averin,Z90,MN,SZ13}.  This property allows to establish a duality relation between Cooper pairs and quantum phase slips.  In particular, the latter can be viewed as effective quantum particles with the topological charge equal to the superconducting flux quantum $\Phi_0=\pi/e$, where $e$ is the electron charge. Such particles tunnel back and forth through the superconducting wire causing not only non-zero average voltage, but also voltage noise \cite{SZ16}. In particular, recently we demonstrated the existence of QPS-induced non-equlibruim shot noise of the voltage in both long and short superconducting
nanowires  \cite{SZ16,SZForsch,SZ17,SZfnt,SZ18}.

In this work we will proceed further and construct a theory describing full counting statistics (FCS) of interacting quantum phase slips in superconducting nanowires. The paper organized as follows. In Sec. \ref{sectS} we describe the system under consideration and define its effective Hamiltonian in the dual representation. In Sec. \ref{sectFCS} we derive the FCS generating function for QPS that allows one to recover all cumulants of the voltage operator in superconducting nanowires. Poissonian nature of the zero-frequency cumulants is demonstrated in Sec. \ref{sectZF}. Sections \ref{sectNP} and \ref{sectSW} are devoted to evaluation of shot noise and higher voltage cumulants at non-zero frequencies.
In Sec. VII we briefly discuss and summarize our key observations. The applicability of our results derived for short superconducting nanowires also to resistively shunted Josephson junctions is demonstrated in Appendix.

\section{The model\label{sectS}}

Below we will address the system depicted in Fig.1. It consists of a superconducting nanowire of length $L$ and cross section $s$ connecting  two big superconducting reservoirs which are in turn attached to external leads. The system is biased by a constant current $I$ and the voltage $V$ across the wire is measured by a detector. As usually, superconductivity inside the wire is described by the fluctuating order parameter field $\Delta (x,t)=|\Delta (x,t)|\exp [i\varphi (x,t)]$, where $t$ denotes real time and $x$ is the coordinate along the wire ($-L/2<x<L/2$).

\begin{figure}
  \includegraphics[width=\columnwidth]{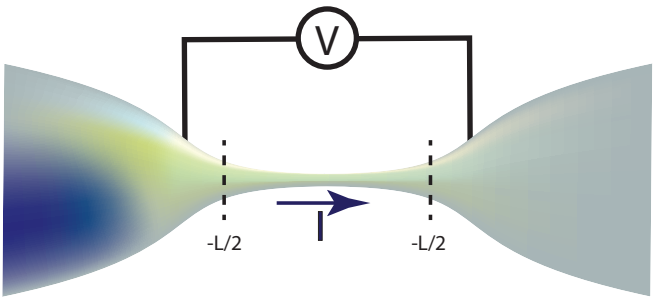}
  \caption {The system under consideration.}  
  \end{figure}

In what follows we will assume that all relevant energy scales, such as, e.g., the frequency $I/e$, temperature $T$ and others remain much smaller than the mean field value $\Delta$ of the order parameter field $|\Delta (x,t)|$ inside the wire. In this case it becomes possible \cite{AGZ,ZGOZ,GZQPS,ogzb} to separate low-energy dynamics of $\Delta (x,t)$  from QPS-related tunneling processes accompanied by local temporary suppression of the order parameter field inside the wire. As before \cite{SZ17,SZfnt,SZ18}, it will be convenient for us to describe the system dynamics in terms of the so-called dual variables $\hat\chi(x)$ and $\hat\Phi(x)$ related to charge density and phase operators $\hat Q$ and $\hat\varphi$ respectively as
\begin{equation}
\hat Q(x)=\frac{1}{\Phi_0}\nabla\hat\chi(x),\qquad \hat\varphi(x)=2e\int\limits_0^x dy\hat\Phi(y).
\end{equation}
These dual field variables obey the standard canonical commutation relation
\begin{equation}
[\hat\Phi(x),\hat\chi(x')]=-i\Phi_0\delta(x-x').
\end{equation}
In the absence of quantum phase slips our superconducting nanowire behaves as a transmission line described by the Hamiltonian
\begin{equation}
\Hat H_{TL} = \int\limits_{-L/2}^{L/2} dx\left(\frac{(\nabla\hat\chi(x))^2}{2\Phi_0^2 C_w}+\frac{\hat\Phi^2(x)}{2\mathcal L_{kin}}\right),
\label{TL}
\end{equation}
where $\mathcal L_{kin}$ is the wire kinetic inductance and $C_w$ is the wire capacitance per unit length. The QPS-related effects are accounted for by the term
\begin{equation}
  \hat H_{QPS}=-\gamma_{QPS}\int\limits_{-L/2}^{L/2} dx \cos\left(\hat \chi(x)\right),
\label{Hqps}
\end{equation}
which follows directly from the commutation relation $[\hat\chi(x),\hat\varphi(x')]=2\pi i\theta(x'-x)$. Here and below
\begin{equation}
\gamma_{QPS} \sim (g_\xi\Delta /\xi)\exp (-ag_\xi), \quad a \sim 1
\end{equation}
is the QPS tunneling amplitude \cite{GZQPS}, $g_\xi =2\pi\sigma_Ns/(e^2\xi) \gg 1$ is the dimensionless normal state conductance of the wire segment of length equal to the superconducting coherence length $\xi$ and $\sigma_N$ is the
Drude conductivity of the wire.

The total Hamiltonian of our system "wire+leads" can be expressed in the form
\begin{equation}
\hat H=\hat H_{TL}+\hat H_{QPS}+\hat H_{env},
\label{totH}
\end{equation}
where the last term $\hat H_{env}$ accounts for an external circuit (environment) as well as for its coupling to the wire.
For simplicity we will assume that both the external environment and its coupling to the wire degrees of freedom are linear. Hence, in the absence of quantum phase slips
(i.e. for  $\gamma_{QPS} \to 0$) the problem remains Gaussian and can be handled exactly. The task at hand is to
include QPS effects into our consideration. As the QPS amplitude always remains sufficiently small this task can
be accomplished by employing a regular perturbation theory in $\gamma_{QPS}$. This approach is appropriate either at not very low energies or, else, on the superconducting side of SIT. The corresponding analysis
is developed below in the next section.

\section{Cumulant generating function \label{sectFCS}}

In order to fully describe voltage fluctuations in the system under consideration it is in general necessary to
evaluate all cumulants of the voltage operator. This goal can be accomplished by deriving the cumulant generating function $\mathcal W$ defined as a logarithm of the so-called "partition function" $\mathcal Z$,
\begin{equation}
  \mathcal W[J]=\ln(\mathcal Z[J])=\ln\left\langle e^{i\int dt J(t)v(t)} \right\rangle,
  \label{cumulantgenf}
\end{equation}
where
\begin{equation}
  v=\frac{1}{\Phi_0C_w}\bigl(\nabla \chi(-L/2)-\nabla \chi(L/2)\bigr)
\end{equation}
is a voltage drop across the wire and $\left\langle ... \right\rangle$ denotes the quantum average fulfilled with the total Hamiltonian  $\hat H$ (\ref{totH}). By taking the $N$-th variational derivatives of $\mathcal W[J]$ with respect to $J(t)$ one
recovers the $N$-th cumulant of the voltage operator (see below).

The function $\mathcal Z[J]$ can be conveniently evaluated by expressing it in terms of a path integral on the Keldysh contour. As usually, all variables of interest are defined on both forward and backward time branches of the Keldysh contour, e.g., $\chi_{F,B}$, giving rise to ``classical'' and ``quantum'' variables, respectively $\chi_+=(\chi_F+\chi_B)/2$ and $\chi_-=\chi_F-\chi_B$ (and similarly for all other operators of interest).  In order to evaluate quantum correlators for any physical quantity it is in general necessary to specify proper time ordering for the corresponding product of operators.  Such ordering becomes insignificant only in the zero frequency limit.  Below in this work we will only be interested in evaluating fully symmetrized cumulants of the voltage operator equivalent to the cumulants of the "classical" variable $v_+(t)$ in our path integral formalism. 
E.g., for $n=2$ we have
\begin{equation}
\langle v_+(t_1)v_+(t_2)\rangle =\frac12\langle\hat v(t_1)\hat v(t_2)+\hat
v(t_2)\hat v(t_1)\rangle,
\end{equation}
while for $n=3$ one finds \cite{GGZ03,SZForsch}
\begin{multline}
\langle v_+(t_1)v_+(t_2)v_+(t_3)\rangle =\frac{1}{8}
  \big\{\langle\hat v(t_1)\big({\cal T}\hat v(t_2)\hat v(t_3)\big)\rangle\\
+\langle\big(\tilde{\cal T}\hat v(t_2)\hat v(t_3)\big)\hat v(t_1)\rangle
+\,\langle\hat v(t_2)\big({\cal T}\hat v(t_1)\hat v(t_3)\big)\rangle\\
+\langle\big(\tilde{\cal T}\hat v(t_1)\hat v(t_3)\big)\hat v(t_2)\rangle
+\,\langle\hat v(t_3)\big({\cal T}\hat v(t_1)\hat v(t_2)\big)\rangle\\
+\langle\big(\tilde{\cal T}\hat v(t_1)\hat v(t_2)\big)\hat v(t_3)\rangle
+\,\langle{\cal T}\hat v(t_1)\hat v(t_2)\hat v(t_3)\rangle\\
+\langle\tilde{\cal T}\hat v(t_1)\hat v(t_2)\hat v(t_3)\rangle\big\},
\label{tri}
\end{multline}
where ${\cal T}$ and $\tilde{\cal T}$ are, respectively, the forward and
backward time ordering operators and $\hat v(t)$ is the voltage drop operator.

With this in mind the function $\mathcal Z[J]$ can be expressed as
\begin{equation}
  \mathcal Z[J]=\left\langle e^{iS_{QPS}[\chi_+,\chi_-]}e^{i\int dt J(t)v_+(t)}\right\rangle_0,
  \label{partf1}
\end{equation}
where \cite{SZ16}
\begin{equation}
  S_{QPS}=-2\gamma_{QPS}\int dt \int\limits_{-L/2}^{L/2} dx\sin(\chi_+)\sin(\chi_-/2),
  \label{SQPS}
\end{equation}
is the action corresponding to the Hamiltonian part (\ref{Hqps}) which accounts for the effect of QPS and  $\langle ... \rangle_0$ denotes averaging with the Gaussian effective action corresponding to the Hamiltonian $\hat H_0=\hat H_{TL}+\hat H_{env}$.
The function (\ref{partf1}) generates voltage correlators
\begin{multline}
\langle v_+(t_1)v_+(t_2)...v_+(t_n)\rangle =\\
\left\langle  v_{+}(t_1)v_{+}(t_2)...
v_{+}(t_n) e^{iS_{QPS}}\right\rangle_0.
\label{VVn}
\end{multline}

In order to proceed let us eliminate the second exponent in Eq. (\ref{partf1}) by making a linear substitution $\chi_{i}=\lambda_{i}+\tilde\chi_i$ and imposing the condition
\begin{equation}
  \left\langle \tilde \chi_i e^{i\int dt J(t)v_+(t)}\right\rangle_0=0
\end{equation}
implying that
\begin{equation}
  \lambda_+(x,t)=\chi_0(x,t)-\int dt'G^K_{\chi v}(x;t,t')J(t'),
\end{equation}
\begin{equation}
  \lambda_-(x,t)=-\int dt' G^A_{\chi v}(x;t,t')J(t').
\end{equation}
Here we denoted $\chi_0\equiv \langle \chi_+ \rangle_0$ and introduced both Keldysh and advanced Green functions (GF), respectively
\begin{equation}
  G^K_{\chi v}(x;t,t')=-i\langle \chi_+(x,t) v_+(t') \rangle_0
\label{KeldGF}
\end{equation}
and
\begin{equation}
  G^A_{\chi v}(x;t,t')=-i\langle \chi_-(x,t) v_+(t') \rangle_0.
\end{equation}
The latter function coincides with the transposed version of the retarded GF
\begin{equation}
  G^R_{\chi v}(x;t,t')=-i\langle \chi_+(x,t) v_-(t') \rangle_0.
\end{equation}

As a result of the above simple manipulations, we obtain
\begin{multline}
  \mathcal Z[J]=e^{-\frac{i}{2}\int dtdt'J(t)G^K_{vv}(t,t')J(t')}\\\times\left\langle e^{iS_{QPS}[\lambda_++\tilde\chi_+,\lambda_-+\tilde\chi_-]}\right\rangle_0,
\end{multline}
where Keldysh GF $G^K_{vv}(t,t')$ is defined analogously to that in Eq. (\ref{KeldGF}). The remaining average can be performed with the aid of the Wick's theorem and expressed via the two GFs,
\begin{eqnarray}
  G^K_{\chi \chi}(x,x';t,t')=-i\langle \tilde \chi_+(x,t) \tilde \chi_+(x',t') \rangle_0,\\
  G^R_{\chi \chi}(x,x';t,t')=-i\langle \tilde \chi_+(x,t) \tilde \chi_-(x',t') \rangle_0,
\end{eqnarray}
while all averages of the type $\langle \tilde\chi_-\tilde\chi_- \rangle_0$ vanish identically due to causality.

Let us now employ the perturbation theory and evaluate the cumulant generating function by expanding $\mathcal Z[J]$ up to the second order in $\gamma_{QPS}$. In this way we get
\begin{multline}
  \mathcal W[J]=-\frac{i}{2}\int dtdt'J(t)G^K_{vv}(t,t')J(t')\\+i\langle S_{QPS}[\lambda_++\tilde\chi_+,\lambda_-+\tilde\chi_-]\rangle_0\\-\frac{1}{2}\langle S^2_{QPS}[\lambda_++\tilde\chi_+,\lambda_-+\tilde\chi_-]\rangle_0\\+\frac{1}{2}\langle S_{QPS}[\lambda_++\tilde\chi_+,\lambda_-+\tilde\chi_-]\rangle_0^2.
\label{W2}
\end{multline}
Substituting the QPS action $S_{QPS}$ (\ref{SQPS}) into Eq. (\ref{W2}) after a simple algebra we observe that the first order contribution in $\gamma_{QPS}$ vanishes and we obtain
\begin{widetext}
\begin{multline}
\mathcal W[J]\approx-\frac{i}{2}\int dtdt'J(t)G^K_{vv}(t,t')J(t')+\gamma_{QPS}^2\int\limits_{-L/2}^{L/2} dxdx'\int dt\int\limits^tdt'\bigl(P(x,x';t,t')-P(x',x;t',t)\bigr)\\\times\sin\biggl(\lambda_{+}(x,t)-\lambda_{+}(x',t')\biggr)\sin\biggl(\frac{\lambda_{-}(x,t)}{2}\biggr)\cos\biggl(\frac{\lambda_{-}(x',t')}{2}\biggr)
  \\-\frac{\gamma_{QPS}^2}{2}\int\limits_{-L/2}^{L/2} dxdx'\int dt\int dt'\bigl(P(x,x';t,t')+P(x',x;t',t)\bigr)\\\times\cos\biggl(\lambda_{+}(x,t)-\lambda_{+}(x',t')\biggr)\sin\biggl(\frac{\lambda_{-}(x,t)}{2}\biggr)\sin\biggl(\frac{\lambda_{-}(x',t')}{2}\biggr),
\label{cumgenf}
\end{multline}
where the function $P(x,x';t,t')$ is defined as \cite{SZ16}
\begin{multline}
  P(x,x';t,t')=\left\langle e^{i\left(\tilde\chi_{+}(x,t)-\tilde\chi_{+}(x',t')-\frac12\tilde\chi_{-}(x,t)-\frac12\tilde\chi_{-}(x',t')\right)}\right\rangle_0=\left\langle e^{i\left(\tilde\chi_{+}(x',t')-\tilde\chi_{+}(x,t)+\frac12\tilde\chi_{-}(x,t)+\frac12\tilde\chi_{-}(x',t')\right)}\right\rangle_0\\=e^{iG^K_{\chi\chi}(x,x';t,t')-\frac{i}2G^K_{\chi\chi}(x,x;t,t)-\frac{i}2G^K_{\chi\chi}(x',x';t',t')+\frac{i}2G^R_{\chi\chi}(x,x';t,t')-\frac{i}2G^A_{\chi\chi}(x,x';t,t')}.
\label{P}
\end{multline}
\end{widetext}

Equation (\ref{cumgenf}) enables one to directly evaluate all (symmetrized) voltage correlators by taking variational derivatives
of $\mathcal W$ with respect to $J(t)$. The structure of this result  actually allows to make an important conclusion even prior to this calculation: It follows immediately from Eq. (\ref{cumgenf}) that in the absence of QPS (i.e. for $\gamma_{QPS} \to 0$) all voltage cumulants except for the second one (describing Gaussian noise of the transmission line (\ref{TL})) vanish identically.  In other words, at low enough temperatures only quantum phase slips give rise to both shot noise of the voltage \cite{SZ16,SZForsch,SZ17,SZfnt,SZ18} and to all higher cumulants of the voltage operator in superconducting nanowires.

We also note that in the considered case of a constant in time current bias $I$ we have $\chi_0(x,t)=I\Phi_0t$ and the function
$P$ depends only on the time difference, i.e. $P(x,x';t,t')=P(x,x';t-t')$. This property will significantly simplify our subsequent
calculations.

\section{Voltage cumulants in the zero frequency limit\label{sectZF}}

To begin with, we employ the above general results in order to evaluate all cumulants of the voltage operator the zero-frequency limit. Proliferation of QPS yields a non-vanishing expectation value $V$ of the voltage operator across our superconducting nanowire \cite{AGZ,ZGOZ,GZQPS} which depends on the external bias current $I$ , i.e. $V=V(I)$.
At the same time an instantaneous voltage value $v(t)$ fluctuates in time due to a sequence of voltage pulses produced by QPS. Let us define the time average
\begin{equation}
  \bar v =\frac{1}{\tau}\int\limits_{-\tau/2}^{\tau/2}d\tau v(\tau),
\end{equation}
with $\tau$ being larger as compared to any relevant time scale for our problem.  It is easy to observe that the cumulants of $\bar v$ are identical to the corresponding cumulants of the voltage operator evaluated in the zero frequency limit. For example, for the first two cumulants one readily finds
\begin{gather}
  \langle\bar v\rangle=\langle v(t)\rangle=V(I),\\
  \langle(\bar v-\langle\bar v\rangle)^2\rangle=\frac{1}{\tau}\int dt\left(\langle v(t) v(0)\rangle-V^2\right)=\frac{1}{\tau}S_0(I).
\end{gather}
Here and below $S_\omega(I)$ denotes the frequency dependent voltage noise power for our wire \cite{SZ16,SZForsch,SZ17,SZfnt,SZ18}.

In order to evaluate the cumulant generating function of $\bar v$
\begin{equation}
  w(j)=\ln\left\langle e^{ij\bar v}\right\rangle
\label{w0}
\end{equation}
it suffices to employ Eq. (\ref{cumgenf}) and set $J(t)=j/\tau$ for $-\tau/2<t<\tau/2$ and $J(t)=0$ otherwise. At large enough values of $\tau$ the combination $\lambda_+(x,t)-\chi_0(x,t)$ becomes practically independent of both $x$ and $t$ implying that $\lambda_{+}(x,t)-\lambda_{+}(x',t')\approx I\Phi_0(t-t')$. Making use of the equation of motion
\begin{equation}
  \biggl(\partial^2_t-\frac{\mathcal L_{kin}}{C_w}\nabla^2\biggr)\hat\chi(x,t)=0
\end{equation}
we conclude that
\begin{equation}
\lim_{\omega\to 0}G^A_{\chi v}(x;\omega)=\lim_{\omega\to 0}G^R_{v\chi}(x;\omega)=\Phi_0
\end{equation}
and, hence, $\lambda_-(x,t)\approx -\Phi_0j/\tau$. As a result we obtain
\begin{multline}
  \frac{w(j)}{\tau}=-\frac{ij^2}{2\tau^2}G^K_{vv}(0)-\frac{\gamma^2_{QPS}}{2}\sin\bigg(\frac{\Phi_0j}{\tau}\bigg)\int\limits_{-L/2}^{L/2} dx dx'\\\times\int\limits_0^\infty dt \bigl(P(x,x';t)-P(x',x;-t)\bigr)\sin(I\Phi_0 t)\\
  -\gamma^2_{QPS}\sin^2\bigg(\frac{\Phi_0 j}{2\tau}\bigg) \int\limits_{-L/2}^{L/2} dx dx' \\\times\int\limits_{0}^\infty dt \bigl(P(x,x';t)+P(x',x;-t)\bigr)\cos(I\Phi_0 t).
\end{multline}
Performing the Fourier transformation
\begin{equation}
  P(x,x';\omega)=\int\limits_0^\infty dt e^{i\omega t}P(x,x';t)
\end{equation}
and defining
\begin{equation}
  \Gamma(\omega)=\frac{\gamma_{QPS}^2}{4}\int\limits_{-L/2}^{L/2}dxdx'\bigl(P(x,x';\omega)+P^*(x',x;\omega)\bigr)
  \label{gamma}
\end{equation}
we cast the above expression for $w$ to a simple form
\begin{multline}
  \frac{w(j)}{\tau}=-\frac{ij^2}{2\tau^2}G^K_{vv}(0)\\+\Gamma(I\Phi_0)\left(e^{\frac{i\Phi_0j}{\tau}}-1\right)+\Gamma(-I\Phi_0)\left(e^{-\frac{i\Phi_0j}{\tau}}-1\right),
\label{wf}
\end{multline}
which fully describes the statistics of QPS-related voltage fluctuations in superconducting nanowires in the zero frequency limit.

It follows immediately from Eq. (\ref{wf}) that this statistics is Poissonian in the above limit \cite{SZForsch}. In particular, combining Eqs. (\ref{w0}) and (\ref{wf}) and evaluating the first and the second derivatives of $w$ with respect to $j$, for the first two voltage cumulants we get
\begin{gather}
\label{fc}  V(I)=\Phi_0(\Gamma(I\Phi_0)-\Gamma(-I\Phi_0)),\\ S_0(I)=iG^K_{vv}(0)+\Phi_0^2(\Gamma(I\Phi_0)+\Gamma(-I\Phi_0)).
\label{sc}
\end{gather}
Equation (\ref{fc}) coincides with the well-known result \cite{ZGOZ} and allows to identify $\Gamma(I\Phi_0)$
as a QPS tunneling rate. Equation (\ref{sc}) reproduces our previous result for the voltage noise \cite{SZ16}.
Employing the detailed balance condition $\Gamma(\omega)=e^{\omega/T}\Gamma(-\omega)$ this result can also
be rewritten as
\begin{equation}
 S_0(I)=iG^K_{vv}(0)+\Phi_0V(I)\coth\left(\frac{I\Phi_0}{2T}\right).
\end{equation}
The two terms in the right-hand side of this formula describe respectively equilibrium Nyquist noise  and QPS-induced shot noise \cite{SZ16}. Note that in the case of the transmission line (\ref{TL}) $G^K_{vv}(0)=0$ and, hence, Nyquist noise vanishes in the zero frequency limit. Nevertheless, here we keep this term for the sake of generality as it can differ from zero in some other models.

Higher voltage cumulants  in the zero frequency limit can be found analogously.  Let us define them as
\begin{equation}
\mathcal C_N(I)=(-i)^N\tau^{N-1}\left.\partial_j^Nw(j)\right|_{j\to0}.
\end{equation}
After a simple algebra all zero frequency cumulants can be expressed through the current-voltage characteristics for our system. In particular, for odd cumulants one has
\begin{equation}
\mathcal C_{2N+1}(I)=\Phi_0^{2N}V(I),
\label{oddc}
\end{equation}
whereas for even ones we obtain
\begin{equation}
\mathcal C_{2N}(I)=\Phi_0^{2N-1}V(I)\coth\left(\frac{I\Phi_0}{2T}\right).
\label{evenc}
\end{equation}
The results derived in this section demonstrate that in the long time limit the effect of interacting QPS  reduces to that of independent sharp voltage pulses which occur with the effective rate $\Gamma(I\Phi_0)$ and are described by Poisson statistics. Note that essentially the same result was previously derived at higher $T$ for thermally activated phase slips (TAPS)\cite{GZTAPS}. At the first sight this similarity can be considered as curious since here we are dealing with quantum interacting objects -- QPS -- which strongly differ from non-interacting classical TAPS. On the other hand, we note that Poisson statistics
for QPS holds only on the superconducting side of SIT where quantum phase slips are bound in pairs which practically do not
interact with each other. With this in mind the similarity between the results derived here and in Ref. \onlinecite{GZTAPS} does not appear very surprising. 

In any case, the above simple physical picture applies only in the zero frequency limit. At non-zero frequencies the system behavior becomes more involved and the statistics of voltage fluctuations deviates from Poissonian, as it will be demonstrated in the next sections.

\section{Noise power in the short wire limit \label{sectNP}}
The general expression for the noise power is defined as
\begin{equation}
  S_\omega(I)=-\int dt e^{i\omega t}\left.\frac{\delta^2\mathcal W[J]}{\delta J(t)\delta J(0)}\right|_{J\to0}.
\end{equation}
With the aid of Eq. (\ref{cumgenf}) we obtain
\begin{widetext}
\begin{multline}
  S_\omega(I)=iG^K_{vv}(\omega)+\frac{\gamma_{QPS}^2}{2}\Biggr[\int\limits_{-L/2}^{L/2}dxdx'G^K_{v\chi}(x;\omega)G^R_{v\chi}(x';\omega)\int\limits_0^\infty
  dt \bigl(P(x,x';t)-P(x',x;-t) \bigr)\cos(I\Phi_0 t)\left(e^{i\omega
      t}-1\right)\\+\frac{1}{4}\int\limits_{-L/2}^{L/2}dxdx'G^R_{v\chi}(x;\omega)G^R_{v\chi}(x';-\omega)\int\limits_{-\infty}^\infty
  dt \bigl(P(x,x';t)+P(x',x;-t) \bigr)\cos(I\Phi_0 t)e^{i\omega t}+\{\omega\to-\omega\}\Biggl].
\label{np}
\end{multline}
\end{widetext}
By virtue of the fluctuation-dissipation theorem this general result can be transformed to that already derived in our previous work \cite{SZ16} where we merely addressed the long wire limit. Here, in contrast, we will specify the expression for the noise power for shorter wires. This limit also covers the case of Josephson junctions and other types of short superconducting contacts.

In order to proceed we observe that each term in the square brackets in Eq. (\ref{np}) contains the combination of the formfactors $P(x,x';\omega)$ describing intrinsic dynamics of a superconducting nanowire during the phase slippage process, as well as two GFs of the $v\chi$-type demonstrating how the detector "feels" voltage fluctuations inside the nanowire. Provided the wire is short enough one can ignore the dependence of these GFs on spatial coordinates and account only for their frequency dependence as
\begin{equation}
G^R_{v\chi}(x;\omega)\approx\Phi_0(1-i\omega\tau_R+...),
\end{equation}
 where $\tau_R$ is the effective RC-time of the system. Accordingly the Keldysh GF can be approximated as
 \begin{equation}
 G^K_{v\chi}(x,\omega)\approx -i\omega\tau_R\coth(\omega/(2T))
 \end{equation}
Employing these approximations, from Eq. (\ref{np}) we obtain
\begin{multline}
  S_{\omega}(I)=iG^K_{vv}(\omega)-i\Phi_0^2\tau_R\omega\coth\left(\frac{\omega}{2T}\right)\bigl(\Gamma^R(\omega+I\Phi_0)\\+\Gamma^R(\omega-I\Phi_0)+\Gamma^R(-\omega+I\Phi_0)+\Gamma^R(-\omega-I\Phi_0)\\-2\Gamma^R(I\Phi_0)-2\Gamma^R(-I\Phi_0)\bigr)+\frac{1}{2}\Phi_0^2\bigl(\Gamma(\omega+I\Phi_0)\\+\Gamma(\omega-I\Phi_0)+\Gamma(-\omega+I\Phi_0)+\Gamma(-\omega-I\Phi_0)\bigr),
\end{multline}
where we introduced the function
\begin{equation}
  \Gamma^R(\omega)=\frac{\gamma_{QPS}^2}{4}\int\limits_{-L/2}^{L/2}dxdx'\bigl(P(x,x';\omega)-P^*(x',x;-\omega)\bigr)
\end{equation}
related to $\Gamma (\omega)$ (\ref{gamma}) by means of the following equation
\begin{equation}
  \Gamma^R(\omega)=\int \frac{dz}{2\pi i}\frac{\Gamma(z)-\Gamma(-z)}{z-\omega-i0}.
\end{equation}

In order to illustrate the above results let us consider a short superconducting nanowire embedded in a linear dissipative external circuit which can, for simplicity, be modeled by an Ohmic shunt resistor $R_S$. As we demonstrate in Appendix,
this situation is equally relevant, e.g. for resistively shunted Josephson junctions in the limit of large Josephson coupling energies $E_J$. In this limit one has
\begin{equation}
  G^R_{\chi\chi}(x,x';\omega)\approx -\frac{2\pi i\mu}{\omega+i0},
\end{equation}
where  $\mu=R_Q/R_S$ is the shunt dimensionless conductance and  $R_Q=\pi/(2e^2)$ is the resistance
quantum.  The QPS tunneling rate then equals to
\begin{equation}
  \Gamma(\omega)=\gamma_{QPS}^2(2\pi T\tau_R)^{2\mu}e^{\frac{\omega}{2T}}\frac{{\bf\Gamma}\left(\mu+\frac{i\omega}{2\pi T}\right){\bf\Gamma}\left(\mu-\frac{i\omega}{2\pi T}\right)}{8\pi T {\bf\Gamma}(2\mu)},
\end{equation}
where ${\bf\Gamma}(y)$ is the Euler gamma-function and $\tau_R^{-1}$ plays the role of effective high-energy cutoff frequency. Evaluating the corresponding integrals in the limit $\omega,T,I\Phi_0\ll \tau_R^{-1}$ and also for $1<\mu<3/2$, we obtain
\begin{equation}
  \Gamma^R(\omega)={\rm const}-i\Gamma(\omega)e^{-\frac{\omega}{2T}}\frac{\sin\left(\pi\mu+\frac{i\omega}{2 T}\right)}{\cos(\pi\mu)}.
\end{equation}
These expressions can be simplified in some limits. For instance, by setting $0<\mu-1\ll 1$ we get 
\begin{equation}
  \Gamma(\omega)\approx\frac{\gamma_{QPS}^2(2\pi T\tau_R)^{2\mu}e^{\frac{\omega}{2T}-2{\bf C}(\mu-1)}\sqrt{\omega^2+4\pi^2 T^2(\mu-1)^2}}{16\pi T^2 {\bf\Gamma}(2\mu)\sqrt{\sin\left(\pi\mu+\frac{i\omega}{2 T}\right)\sin\left(\pi\mu+\frac{i\omega}{2 T}\right)}},
\end{equation}
where $\bf C$ is Euler-Mascheroni constant. Also the expressions for the QPS tunneling rate are simplified greatly for $|\omega|\gg T$. One has
\begin{equation}
\Gamma(\omega)\approx\pi\gamma_{QPS}^2\theta(\omega)\frac{(\omega\tau_R)^{2\mu}}{2\omega\Gamma(2\mu)},
\end{equation}
\begin{equation}
\Gamma^R(\omega)\approx{\rm const}+\pi\gamma_{QPS}^2\frac{|\omega\tau_R|^{2\mu}e^{-i\pi\mu{\rm sign}(\omega)}}{4\omega\Gamma(2\mu)\cos(\pi\mu)}.
\end{equation}
Accordingly, in the zero-temperature limit one finds
\begin{equation}
\mathcal C_N(I)=\pi\gamma_{QPS}^2{\rm sign}^N(I)\frac{\Phi_0^{N+2\mu-1}\tau_R^{2\mu}}{2\Gamma(2\mu)}|I|^{2\mu-1}.
\end{equation}
Note that the above results are consistent with ones derived in \cite{ANO}.

\section{Higher voltage cumulants \label{sectSW}}
Let us now turn to higher voltage cumulants at non-zero frequencies. It is instructive to define a general expression for the frequency dependent $N$-th voltage cumulant as
\begin{multline}
  S_{\omega_1,...,\omega_{N-1}}(I)=\int dt_1...dt_{N-1} e^{i\omega_1t_1+...+i\omega_{N-1}t_{N-1}}\\\times(-i)^N\left.\frac{\delta^N \mathcal W[J]}{\delta J(t_{N-1})...\delta J(t_1)\delta J(0)}\right|_{J\to 0}.
\end{multline}
Note that from definition it follows that
\begin{equation}
S_{\underbrace{00...0}_N}(I)=\mathcal C_N(I).
\end{equation}
In the limit $T,\omega\ll \tau_R^{-1}$ or, in other words, provided the detector immediately "feels" voltage fluctuations
generated by quantum phase slips, one can set $\tau_R\to 0$ and explicitly evaluate all voltage cumulants at non-zero frequencies. In this case for the cumulant generating function we get
\begin{multline}
  \mathcal W[J]\approx-\frac{i}{2}\int dtdt'J(t)G^K_{vv}(t-t')J(t')\\-\gamma_{QPS}^2\int\limits_{-L/2}^{L/2} dxdx'\int dt\int\limits^tdt'\bigl(P(x,x';t-t')-P(x',x;t'-t)\bigr)\\\times\sin\bigl(I\Phi_0(t-t')\bigr)\sin\biggl(\frac{\Phi_0J(t)}{2}\biggr)\cos\biggl(\frac{\Phi_0 J(t')}{2}\biggr)
  \\-\frac{\gamma_{QPS}^2}{2}\int\limits_{-L/2}^{L/2} dxdx'\int dt\int dt'\bigl(P(x,x';t-t')+P(x',x;t'-t)\bigr)\\\times\cos\bigl(I\Phi_0 (t-t')\bigr)\sin\biggl(\frac{\Phi_0 J(t)}{2}\biggr)\sin\biggl(\frac{\Phi_0J(t')}{2}\biggr).
  \label{cumgenfs}
\end{multline}

In is straightforward to observe that the second term in Eq. (\ref{cumgenfs}) can only contribute to odd cumulants, whereas the last term, in contrast, determines all even cumulants. After some algebra we arrive at the following expressions for both even and odd voltage cumulants, respectively
\begin{multline}
  S_{\omega_{1},...,\omega_{2M}}(I)=\frac{\Phi_0^{2M+1}}{2^{2M}(2M)!}\sum_{p\in{\rm perm}} \sum_{m=0}^{2M} \binom{2M}{m} \\ \times\Bigl(\Gamma^R\bigl(I\Phi_0-(-1)^m(\omega_{p_1}+...+\omega_{p_m})\bigr)\\-\Gamma^R\bigl(-I\Phi_0-(-1)^m(\omega_{p_1}+...+\omega_{p_m})\bigr)\Bigr)
\end{multline}
and
\begin{multline}
  S_{\omega_{1},...,\omega_{2M+1}}(I)\\=\frac{\Phi_0^{2M+2}}{2^{2M+1}(2M+1)!}\sum_{p\in{\rm perm}} \sum_{m=0}^{M} \binom{2M+1}{2m+1}\\ \times\Bigl(\Gamma\bigl(I\Phi_0+(\omega_{p_1}+...+\omega_{p_{2m+1}})\bigr) \\+\Gamma\bigl(-I\Phi_0+(\omega_{p_1}+...+\omega_{p_{2m+1}})\bigr)
  \\+\Gamma\bigl(I\Phi_0-(\omega_{p_1}+...+\omega_{p_{2m+1}})\bigr)
  \\+\Gamma\bigl(-I\Phi_0-(\omega_{p_1}+...+\omega_{p_{2m+1}})\bigr)
  \Bigr).
\end{multline}
Here the sum is taken over all permutations of frequencies.

The above results allow one one to extend the relation between the voltage cumulants and the current-voltage characteristics of our device to non-zero frequencies. For the odd cumulants one finds
\begin{widetext}
\begin{equation}
  S_{\omega_{1},...,\omega_{2M}}(I)=\frac{\Phi_0^{2M+2}I}{2^{2M-1}(2M)!}\int \frac{dI'}{2\pi i} V(I') \sum_{p\in{\rm perm}} \sum_{m=0}^{2M} \binom{2M}{m}
  \frac{1}{(I'\Phi_0+(-1)^m(\omega_{p_1}+...+\omega_{p_m})-i0)^2-(I\Phi_0)^2},
\label{evencw}
\end{equation}
whereas the expression for the even cumulants reads
\begin{multline}
  S_{\omega_{1},...,\omega_{2M+1}}(I)=\frac{\Phi_0^{2M+1}}{2^{2M+1}(2M+1)!}\sum_{p\in{\rm perm}} \sum_{m=0}^{M} \binom{2M+1}{2m+1}\\\times\Biggl(\coth\biggl(\frac{I\Phi_0+(\omega_{p_1}+...+\omega_{p_{2m+1}})}{2T}\biggr)V\biggl(I+\frac{\omega_{p_1}+...+\omega_{p_{2m+1}}}{\Phi_0}\biggr)
  \\+\coth\biggl(\frac{I\Phi_0-(\omega_{p_1}+...+\omega_{p_{2m+1}})}{2T}\biggr)V\biggl(I-\frac{\omega_{p_1}+...+\omega_{p_{2m+1}}}{\Phi_0}\biggr)
  \Biggr).
  \label{oddcw}
  \end{multline}
\end{widetext}

\begin{figure}
  \includegraphics[width=\columnwidth]{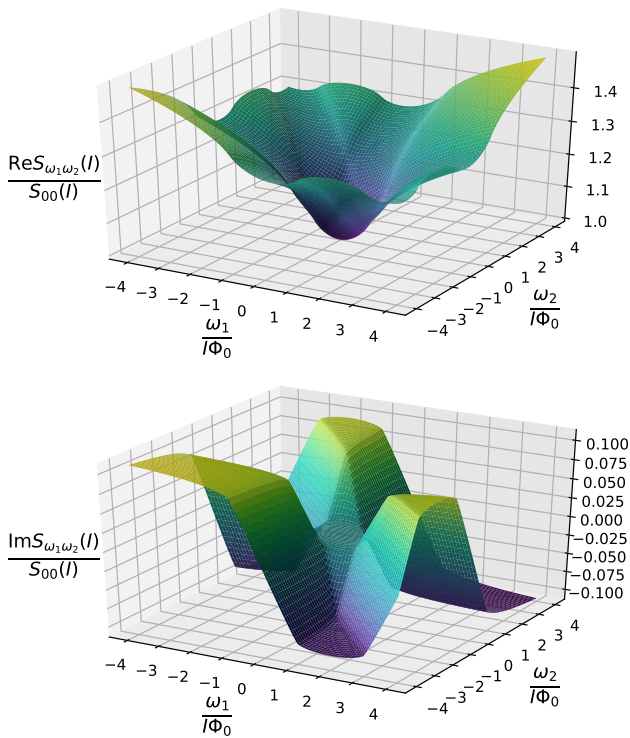}
\caption{Real and imaginary parts of the third voltage cumulant at $T \to 0$ and $\mu=1.1$.}
\label{figres1}
\end{figure}

\begin{figure}
  \includegraphics[width=\columnwidth]{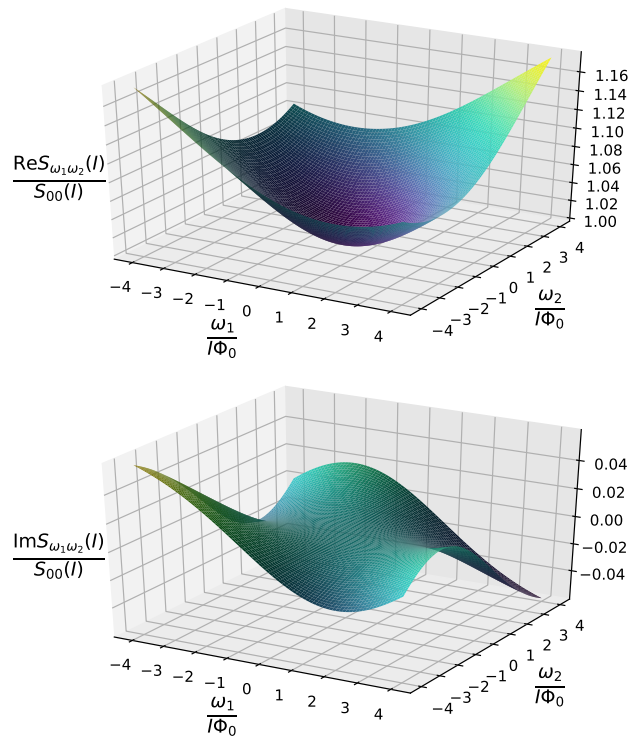}
\caption{The same as in Fig. \ref{figres1} at $T=I\Phi_0$.}
\label{figres2}
\end{figure}

These expressions can be evaluated numerically. The corresponding results for the third voltage cumulant as a function of two frequencies are displayed in Figs. \ref{figres1} and \ref{figres2} respectively in the limits of low and high temperatures. We observe that the third voltage cumulant consists of real and imaginary parts 
\begin{equation}
S_{\omega_{1},\omega_{2}}(I)={\rm Re}S_{\omega_{1},\omega_{2}}(I)+i {\rm Im}S_{\omega_{1},\omega_{2}}(I).
\end{equation}
Both these functions become considerably smoother at higher $T$.


\section{Discussion and conclusions}
In this work we developed a microscopic theory enabling one to fully describe statistics of voltage fluctuations generated by quantum phase slips in superconducting nanowires. For this purpose we evaluated the cumulant generating function that contains complete information about all voltage correlators in such nanowires. Already from the form of this function it is easy to observe that the third and all higher voltage cumulants differ from zero only due to the presence of QPS and vanish identically should the effect of quantum phase slips be neglected. Likewise, quantum phase slips are responsible for the presence of shot noise of the voltage in superconducting nanowires  \cite{SZ16,SZForsch,SZ17,SZfnt,SZ18}.

Note, that previously various aspects of fluctuation statistics have been  addressed by a number of authors in the case of normal mesoscopic conductors (see, e.g., Refs. \onlinecite{GGZ03,Nag2003,KNB2004} and further references therein) as well as for superconducting structures, such as quasi-one-dimensional wires \cite{GZTAPS} and resistively shunted Josephson junctions \cite{GMU2010, ZBN2015}. It is worth pointing out that the authors  \cite{GZTAPS,GMU2010, ZBN2015} restricted their analysis to thermal fluctuations and, hence, their results remain applicable at not too low temperatures.
Here, in contrast, we set up a fully quantum mechanical treatment of the problem that essentially operates with interacting quantum phase slips and allows to fully describe statistics of voltage fluctuations at any temperature down to $T \to 0$.

Proceeding perturbatively in the QPS tunneling rate we demonstrated that at long times or, equivalently, in the zero-frequency limit the statistics of voltage fluctuations in superconducting nanowires reduces to Poissonian one similarly to the situation encountered in a number of other tunneling-like problems. Furthermore, it is straightforward to show that in this limit all (symmeterized) cumulants of the voltage operator can be expressed in a simple manner through the current-voltage characteristics of the system $V(I)$, cf. Eqs. (\ref{oddc}),  (\ref{evenc}). At non-zero frequencies, however, quantum voltage fluctuations in superconducting nanowires are not anymore described by Poisson statistics. This is because inter-QPS interaction produced by an effective environment (due to the wire itself and/or an external dissipative circuit) starts playing
a more important role at shorter time scales making the whole problem much more involved.  Remarkably, also in this case it is possible to establish a relation between the voltage cumulants and the current-voltage characteristics of our device $V(I)$, though in a much more complicated form as compared to that in the zero frequency limit, cf. Eqs.  (\ref{evencw}),  (\ref{oddcw}).
The latter observation could be important for possible experimental verification of our theoretical predictions.

\vspace{0.5cm}

\centerline{\bf Acknowledgements}
This work is a part of joint Russian-Greek Projects
No. RFMEFI61717X0001 and No. T4$\Delta$P$\Omega$-00031 "Experimental
and theoretical studies of physical properties of low-dimensional
quantum nanoelectronic systems".

\appendix

\section{}

Consider a current-biased resistively shunted Josephson junction described by the 
Hamiltonian \cite{SZ90}
\begin{equation}
\hat H = \frac{\hat Q^2}{2C_J}-E_J\cos(\hat \varphi)-\frac{I\hat\varphi}{2e}+\hat H_{R}[\hat\varphi],
\end{equation}
where $C_J$ and $E_J$ are, respectively, the junction capacitance and the Josephson coupling energy. 
The charge and the phase operators $\hat Q$ and $\hat\varphi$ obey the standard commutation relation $[\hat Q,\hat\varphi]=-2ie$. Finally, the term $\hat H_R$ accounts for an external resistor which can be routinely described, e.g., within the standard Caldeira-Leggett model. As before, the voltage operator is defined as $\hat
V=\partial_t\hat\varphi/(2e)=\hat Q/C_J$. 

In the limit of large $E_J \gg E_C=e^2/2C_J$ the junction phase dynamics is determined
by quantum tunneling between the minima of the cosine potential $-E_J\cos\varphi$. Let us derive an
effective low-energy Hamiltonian for our junction in the limit of large $E_J$. For this purpose it will be convenient for us to extend the Hilbert space for our system by introducing an extra pair of canonically conjugated variables $\hat \phi,\ \hat q$ obeying the commutation relation $[\hat q,\hat\phi]=-2ie$.  Consider the Hamiltonian
\begin{equation}
  \hat H_{ext}= \frac{(\hat Q+\hat q)^2}{2C_J}-E_J\cos(\hat\phi)-\frac{I\hat\varphi}{2e}+\hat H_{R}[\hat\varphi],
\end{equation}
where the variable $\phi$ is treated as compact implying that
the eigenvalues of $\hat q$ are proportional to integer numbers. It is straightforward to observe that
the operator $\hat\phi-\hat\varphi$ commutes with the Hamiltonian. Hence, the whole
Hilbert space for our system can be split into subspaces with fixed values of
$\hat\phi-\hat\varphi$ and the system dynamics described by the extended Hamiltonian  $\hat H_{ext}$ coincides with that governed by the initial Hamiltonian within the subspace $\hat\phi=\hat\varphi \mod 2\pi$. 

Now let us recall that in the limit $E_J\to\infty$ the variables $\hat q,\hat \phi$ can be treated as fast ones in contrast to $\hat Q,\hat \varphi$ which represent slow variables. Let us trace out the two fast variables and then build up a special basis by introducing the eigenvectors $\hat Q|Q\rangle=Q|Q\rangle$ and
\begin{equation}
\left(\frac{(\hat q+Q_x)^2}{2C_J}-E_J\cos(\hat\phi)\right)|n,Q_x\rangle = E_n(Q_x)|n,Q_x\rangle
\end{equation}
The last vector is just the  Bloch state. In the limit of large $E_J$ only the value $n=0$ matter and, hence, one can project the Hamiltonian onto the corresponding subspace. This procedure is performed with the aid of the projector 
\begin{equation}
   \mathcal P=\int dQ |Q\rangle\langle Q|\otimes|0,Q\rangle\langle 0,Q|.
\end{equation}
As a result we obtain
\begin{equation}
 \hat H_{JJ} = E_0(\hat Q)-\frac{I\hat\varphi}{2e}+\hat H_{R}[\hat\varphi]
\end{equation}
Making use of translational invariance we conclude that the energy $E(Q_x)$ is a $2e$-periodic function of the charge $Q_x$. Introducing new variables $\hat \Phi=\hat\varphi/(2e)$, $\hat
\chi=-\pi\hat Q/e$ and setting $E_0(Q_x)\approx -\gamma\cos(\pi Q_x/e)$ we arrive at the effective Hamiltonian 
\begin{equation}
\hat H_{JJ,sc}=-I\hat\Phi-\gamma\cos(\hat\chi)+\hat H_{R}[2e\hat\Phi]
\end{equation}
very similar to $\hat H_{TL}+\hat H_{QPS}$ if one neglects the spatial dependence of $\hat\chi(x)$ and $\hat\Phi(x)$. Hence, all our results derived here for short superconducting nanowires can equally be applied to Josephson junctions in the limit of large $E_J$ by replacing $\gamma_{QPS}\to\gamma$ and formally considering the proper $G^K_{\chi\chi}$ corresponding to an external bath described by the Hamiltonian $\hat H_{R}[2e\hat\Phi]$.



\begin{thebibliography}{30}
\bibitem{LV} A.I. Larkin and A.A. Varlamov, {\it Theory of fluctuations in superconductors} (Clarendon, Oxford, 2005).
\bibitem{AGZ} K.Yu. Arutyunov, D.S. Golubev, and A.D. Zaikin, Phys. Rep. {\bf 464}, 1 (2008).
\bibitem{ms} J.E. Mooij and G. Sch\"{o}n, Phys. Rev. Lett. {\bf 55}, 114 (1985).
\bibitem{Bu} B. Camarota, F. Parage, F. Balestro, P. Delsing, and O. Buisson, Phys. Rev. Lett. {\bf 86}, 480 (2001).
\bibitem{RSZ} A.A. Radkevich, A.G. Semenov, and A.D. Zaikin, Phys. Rev. B {\bf 96}, 085435 (2017).
\bibitem{ALRSZ} K.Yu. Arutyunov, J.S. Lehtinen, A.A. Radkevich, A.G. Semenov, and A.D. Zaikin, J. Magn. Magn. Mat. {\bf 459}, 356 (2018).
\bibitem{ZGOZ} A.D. Zaikin, D.S. Golubev, A. van Otterlo, and G.T. Zimanyi, Phys. Rev. Lett. {\bf 78}, 1552 (1997).
\bibitem{GZQPS} D.S. Golubev and A.D. Zaikin, Phys. Rev. B {\bf 64}, 014504 (2001).
\bibitem{BT} A. Bezryadin, C.N. Lau, and M. Tinkham, Nature {\bf 404}, 971 (2000).
\bibitem{Lau} C.N. Lau, N. Markovic, M. Bockrath, A. Bezryadin, and M. Tinkham, Phys. Rev. Lett. {\bf 87},  217003 (2001).
\bibitem{Zgi08} M. Zgirski, K.P. Riikonen, V. Touboltsev, and K.Y. Arutyunov, Phys. Rev. B {\bf 77},  054508 (2008).
\bibitem{AstNature} O.V. Astafiev, L.B. Ioffe, S. Kafanov, Yu.A. Pashkin, K.Yu. Arutyunov, D. Shahar, O. Cohen, and J.S. Tsai, Nature {\bf 484}, 355 (2012).
\bibitem{PZ88} S.V. Panyukov and A.D. Zaikin, J. Low Temp. Phys. {\bf 73}, 1 (1988).
\bibitem{averin} D.V. Averin and A.A. Odintsov, Phys. Lett. A {\bf 140}, 251 (1989).
\bibitem{Z90} A.D. Zaikin, J. Low Temp. Phys. {\bf 80}, 223 (1990).
\bibitem{MN} J.E. Mooij and Yu.V. Nazarov, Nat. Phys. {\bf 2}, 169 (2006).
\bibitem{SZ13} A.G. Semenov and A.D. Zaikin, Phys. Rev. B {\bf 88}, 054505 (2013).
\bibitem{SZ16} A.G. Semenov and A.D. Zaikin, Phys. Rev. B {\bf 94}, 014512 (2016).
\bibitem{SZForsch} A.G. Semenov and A.D. Zaikin, Fortschr. Phys.  1600043 (2017).
\bibitem{SZ17} A.G. Semenov and A.D. Zaikin, J. Supercond. Nov. Magn. {\bf 30}, 139 (2017).
\bibitem{SZfnt} A.G. Semenov and A.D. Zaikin, Fiz. Nizk. Temp. (Kharkov) {\bf 42}, 1011 (2017).
\bibitem{SZ18} A.G. Semenov and A.D. Zaikin, J. Supercond. Nov. Magn. {\bf 31}, 711 (2018).
\bibitem{ogzb} A. van Otterlo, D.S. Golubev, A.D. Zaikin, and G. Blatter, Eur. Phys. J. B {\bf 10}, 131 (1999).
\bibitem{GGZ03}  A.V. Galaktionov, D.S. Golubev, and A.D. Zaikin, Phys. Rev. B {\bf 68}, 235333 (2003).
\bibitem{GZTAPS} D.S. Golubev and A.D. Zaikin, Phys. Rev. B {\bf 78}, 144502 (2008).
\bibitem{ANO} D.V. Averin, Yu.V. Nazarov, and A.A. Odintsov, Physica B {\bf 165-166}, 945 (1990).
\bibitem{Nag2003} K.E. Nagaev, cond-mat/0302008 (unpublished).
\bibitem{KNB2004} M. Kindermann, Yu. V. Nazarov, and C.W.J. Beenakker, Phys. Rev. B {\bf 69}, 035336 (2004).
\bibitem{GMU2010} D.S. Golubev, M. Marthaler, Y. Utsumi and G. Sch\"on, Phys. Rev. B {\bf 81}, 184516 (2010).
\bibitem{ZBN2015} M. Zonda, W. Belzig and T. Novotny, Phys. Rev. B {\bf 91}, 134305 (2015).
\bibitem{SZ90} G. Sch\"on and A.D. Zaikin, Phys. Rep. {\bf 198}, 237 (1990).
\end{thebibliography}
\end{document}